# ON SOME DETAILED EXAMPLES OF QUANTUM LIKE STRUCTURES CONTAINING QUANTUM POTENTIAL STATES ACTING IN THE SPHERE OF BIOLOGICAL DYNAMICS.

Elio Conte [1], Gianpaolo Pierri [1], Leonardo Mendolicchio[1], Andrei Yu. Khrennikov [2], Joseph P. Zbilut [3]

[1] Department of Pharmacology and Human Physiology, TIRES-Center for Innovative Technologies for Signal Detection and Processing, and Department of Neurological Sciences, Unit of Psychiatry, University of Bari, Policlinico, I-70124, Bari, Italy
[2] International Center for Mathematical Modeling in Physics, Engineering, Economy and Cognitive Sciences, University of Vaxjo, Sweden
[3] Department of Physiology, Rush Medical College, Rush University, Chicago, Illinois, 60612 USA

**Abstract**: In the first part of the present paper we give an analysis of the ontic nature of quantum states to be intended as potentialities and of the central role of spin to be considered as the basic essence of quantum mechanical reality: using an algebraic quantum like structure we give mathematical proof on the transition from potentiality to actualization : we recall here what was recently given by us in arXiv quant-ph/0607196. However, as may be expected, it is not so easy to introduce examples containing an adequate description of ontic potentialities through detailed models of systems. The central aim of this paper is to attempt to reach this objective giving direct cases of systems in which ontic potentialities act jointly to actualization. Our aim is to provide evidence for the possible importance of potential states in the sphere of the biological dynamics giving detailed examples of interest for biological studies. We outline the possible implications of potentialities at the level of linear and non linear biological dynamics.

## 1. Introduction

It is known that the problem of how a mathematical superposition of manifold possibilities evolves to become a particular observable actuality, represents the basic unsolved problem of measurement in quantum mechanics. In some sense it delineates also a basic unsolved question of our science and knowledge. In fact, the present quantum problem may be considered the last more modern version of a problematicism and a debate that involved our science and philosophy from the past, starting with Empedocle, Plato and the same Aristotle who first considered that potentia and actuality are two kinds of reality and that actualities give origin to potentia which give origin to actualities. W. Heisenberg evidenced the value of quantum mechanics in this Aristotelian basic principle. On the other hand, quantum theory is not able to reach adequate evidence on the nature of such potential entities, on their qualification between their quantum actualizations and, finally, on the same mathematical and physical features regulating the transition in our reality from potential to actual entities. The aim of the present paper is to move in the framework of a quantum like formulation, giving for the first time a mathematical proof of such possible transition from potential to actual entities in our reality and simultaneously describing the mathematical and physical features that characterize such transition.

Before to proceed with the mathematical proof, we need to deepen two important questions. The first relates the nature of the quantum states and the second considers the nature of the spin. We retain that this last quantum observable represents the essence of quantum mechanics and therefore our proof will be based on the utilization at the algebraic level of basic abstract elements that, as we shall see, move in direct analogy with the notion of spin.

**2. Some Observations on the Nature of Quantum States**.
Let us start with a preliminary analysis on the notion of quantum state. The problem here is well known. Let us consider a particle, as example an electron, impinging on a screen. According to quantum mechanics we cannot know where it will hit but we can always assign probabilities to potentialities of the electron to hit at different locations. These are given by the well known wave function or quantum state of the system we have in consideration. At some time, the electron impinges into some point of the screen and since it hits the screen, we have no more a matter of probability. The quantum state has collapsed into some definite point on the screen. The arising question is on the nature of the quantum state. In an epistemic interpretation we consider that the quantum state describes not the system in consideration but our status of knowledge about it. In an ontic interpretation we admit instead that the quantum states are ontic and this is to say that they describe the system as it is. The point here is to consider an ontic nature of the quantum state but the settlement of its definition is paved with conceptual difficulties. Some possibilities to proceed with an ontic interpretation of quantum states were previously explored [1]. However, undoubtedly, it is not so easy to introduce an adequate notion of ontic potentialities in our reality. As example, it seems rather an approximation for defect to consider here a superposition state of potentialities meaning that the system can be in two or more states at the same time. In fact, we must remember here that the entity in consideration is a potential and not an actual entity. It must be considered to be real, ontologically significant, but not being actual. By the previous definition we ran the risk to consider the coexistence of potentialities as an actual like form, that is a superposition of coexisting like actualities and this is not what the quantum superposition principle admits. This is one first difficulty. One other question arises in the following manner. Let us consider two quantum non commuting entities A and B. Quantum mechanics tells us that, if one such entity, say A, is actualized, B consequently remains an undefined potentiality. In our opinion, this is an ontic holistic process that must receive a proper general, mathematical formalization while instead in this case the traditional quantum formalism, based on the Hilbert space formulation, holds only the requirement of mutually orthogonal vectors that are representative of the mutual exclusivity of the states. The consequent mutual exclusivity of outcome states is merely an epistemic phenomenon and thus ontologically insignificant while instead we need in this case an holistic description having a full ontological explanation. This is one of the reasons because we introduce in the following section a quantum like schema, not based as traditionally on quantum linear operators and Hilbert spaces, but fixed instead on an algebraic structure and its formalization and where the auspicated ontological significance of the holistic process regarding at the same time the actualization of the entity A and the persisting potentiality of the entity B, is reached through the proof of the theorem 2. Here the notion of holism linking actuality to potentiality seems absolutely necessary. In conclusion, we like more to steer ourselves into a different definition [see also 1] considering an ontic superposition of potentialities meaning an ontic holistic entanglement where no more independently existing features of potentiality as actual like forms may be identified. In addition, linked coexisting forms of actuality and potentiality must be expected in our reality still in a whole holistic ontological framework. We would also give some evidence about such definitions. Let us admit hypothetically an existing system S that may be represented by an actual state, that we identify by $\psi_{n,A}$, and a potential state that is given in a multiplicative manner by $\psi_{n,P} = \varphi_{n,P} X_P$ where here $X_P$ is the potential entity, a symbol whose only quantum constraint is to be $X_P^2 = 1$ so that its potentiality is to be actualized as +1 or -1. $\varphi_{n,P}$ represents instead some scalar quantity connected to $X_P$. The time evolution of such system is admitted to be given by the following map

$$\psi_{n+1,A} + \varphi_{n+1,P}X_P = (\psi_{n,A} + \varphi_{n,P}X_P)(\psi_{n,A} + \varphi_{n,P}X_P) = \psi_{n,A}^2 + \varphi_{n,P}^2 + 2\psi_{n,A}\varphi_{n,A}X_P$$

In the evolution of such hypothetical system, we have an actualized entity, $\psi_{n,A}^2 + \varphi_{n,P}^2$, which at any stage of the evolution experiences both the actual ($\psi_{n,A}^2$) and the potential ($\varphi_{n,P}^2$) contributions. Instead, the potential entity is represented by the term $2\psi_{n,A}\varphi_{n,A}X_P$, where the potential symbol, $X_P$, goes on maintaining at each step, its potentiality to be +1 and -1. Finally, at some time $X_P$ may be considered to randomly assume an actualized value of +1 or of -1 giving in this case a final actualized value for the whole evolution process in consideration. This is an example of potential – actualized process that at last in principle is based on a quantum like scheme. As we see, it changes radically our traditional view on dynamics of reality. The time evolution of the system starts with the actualized value $\psi_{0,A}$. However, with basic difference respect to our traditional view on evolution processes, it has also the potentiality to be or $\psi_{0,A} + \varphi_{0,P}$ or $\psi_{0,A} - \varphi_{0,P}$. This is an intrinsic potentiality of such evolution process. In the future steps of the evolution the system maintains its potentialities that at each stage will be given directly by the basic mathematical features of the map. An occurring actualization of the process randomly at some time will attribute to $X_P$ a definite value, or +1 or -1, and this actualization will enable the evolution process to actualize a final value of the evolutive process that will account also of the previously unexpressed (unactualized) potentialities.

Consider another scheme that may be of basic importance in studies on biological dynamics and in the non linear analysis of systems.

Let us assume the following logistic map

$$x_{n+1} = kx_n(x_n - 1)$$

It is known that it represents one of the most investigated deterministic chaotic maps. Its properties of chaotic deterministic behaviour for $3.7 \leq k \leq 4.0$ are well known and well established [2, soon after increase reference number of one]. We may extend such map in a quantum like scheme so to give a logistic map with included potential quantum like features. In this case the usual logistic map becomes

$$A_{n+1} + B_{n+1}X_P = k(A_n + B_nX_P)(A_n + B_nX_P - 1)$$

As previously discussed, $X_P$ represents the potential entity. We have $X_P^2 = 1$ and by actualization at some time it may assume a definite numerical value of +1 or of −1. The algebraic principle of identities for algebraic quantities gives the following two maps

$$A_{n+1} = kA_n^2 + kB_n^2 - kA_n$$

and

$$B_{n+1}X_P = (kA_nB_n - kB_n)X_P$$

The first map pertains to the actualized component, $A_n$, of such quantum like logistic map. The basic feature is that, at each stage of its evolution, such actualized component experiences the influence of the potential component by the presence of $B_n^2$ in each stage of its evolution. The second map pertains instead to the potential component of the given quantum like logistic map, $B_n$. At each stage of the iteration such potential component also experiences the presence of the actualized component by $A_n$. We see that our traditional manner to look at the processes of our reality, changes radically. Now, potential and actual entities cooperate to give the time dynamics of the considered process. As a model we introduce basic quantum like systems and mathematically, we propose to consider dynamical systems in Clifford algebras. We will explore in detail such new non linear phenomenology in following papers.

We may also examine more articulated quantum schemes in which the potential contribution, previously expressed by $\varphi_{n,P} X_P$, will be now replaced by a more general term of the form $F(X_{1,P}, X_{2,P}, X_{3,P})$ where the $X_{i,P}$ are this time three potential symbols given as example as in the (11) of the following section and realizing in this manner a quantum like scheme in which also the contributions of the non commutativity are taken in consideration.

In conclusion, we retain that quantum potentialities, as roughly expressed by the previous model, find their principal arena in the sphere of the biological matter. In the last section of this paper, we will give some detailed examples of biological themes in which symbolic potential entities as $X_{i,P}$ may be involved. Potentialities may explain their decisive role when discussing the basic theme of the Neodarwinism that of course was just the object of a recent investigation in the framework of quantum potentialities [2 to be increased by one]. In this case what radically changes respect to our traditional manner to conceive evolution is that in classical Neodarwinism we have evolution essentially intended as consequence of random variations and natural selection of what is the fittest form. Here we have selection of forms of concrete and actual matter. Instead in the case of the evolution model previously introduced and having potentialities, we have similarly a final selection of concrete and actual forms but the arena of the possible differentiation is extremely different and, in particular, forms of potentiality this time coexist with forms of actuality, and potentiality contributes to characterize actual forms at each stage of the evolution. It is sufficient to look at the previously given relation to convince that we are in fact in presence of a radically new kind of evolution mechanism where, we repeat, forms of potentiality coexist with forms of actuality. The result is in a new structure which makes possible many more possible and different pathways that result impossible in an evolution mechanism based only on actualization. At the same time it is the basic concept of reality that changes radically in the sense that in the traditional case it is the concrete and actual matter that constitutes the ontological reference of a basic materialistic instance while instead in such new case matter must be considered in its potential form to be envisaged in addition to its actual form.

**3. Observations on the Nature of Spin.**
W. Pauli was the physicist that had a decisive role in the elaboration of the quantum theory of spin. Initially, he called the spin a "two-valued quantum degree of freedom". On the basis of this definition, we retain that initially he considered the spin as a kind of physical-logical -informative entity linked to matter at the microphysical level. His definition of spin remained initially rather vague and uncharacterized until R. Kroning in 1925 suggested that it would be produced by the self-rotation of the electron. This was an idea that Pauli initially criticized severely but Kroning's view on spin was subsequently supported from G.Uhlenbeck and S. Goudsmit in the same year, and finally Pauli, despite his initial objections to this idea, formalized the theory of spin in 1927, accepting to characterize it as self-rotation of a quantum particle. He pioneered the use of the so called Pauli matrices as a representation of spin operators, and he introduced a two-component spinor wave function. His spin-theory was not relativistic. In 1928 P. Dirac described the relativistic electron with a four component spinor, and he found that the spin is a relativistic effect that may be identified by linearization of the Hamiltonian in special relativity [3].

At the present we are convinced that the spin represents an entity of Nature whose meaning and role are more general, indeed universal, with respect to the rather restrictive interpretation that was originally formulated by Kroning, Uhlenbeck and Goudsmit. We retain that the following examples are of basic importance to accept our thesis. Quantum computing has introduced the qubit quantum register. Here a universal unity of information is formulated in quantum mechanical terms and information no longer results a rather abstract entity but for the first time it is really and tangiblely connected to a material object as an elementary particle. In addition, P. O'Hara [4] obtained that the spin is introduced in a natural way into the space-time metric, taking the square root of the metric associated with space Also other authors, as P. Cordero, C. Teitelboim, and R. Tabensky [5] introduced the spin into relativity but taking the square root of the Hamiltonian. As well as in the Dirac equation the spin was obtained by linearizing the Hamiltonian of special relativity, in [4] spin matrices were obtained by linearizing not the Hamiltonian of relativity but rather the space-metric itself, and this result provided the conclusion in [4] that the spin is intrinsically linked to the geometrical properties of space-time. Such results indicate that the so called spin must be really intended as manifestation of a general and universal entity having in Nature an articulated, physical, informative, and logic role. Other interesting evidences of such a conclusion may be reached through investigation in biology and physiology. Two authors, Hu. Hu and M. Wu, [6], introduced a theory in which the advent of consciousness is intrinsically connected to spin. They formulated a spin-mediated consciousness theory based on pan-protopsychism. These authors were able to discuss a well defined neurophysiological model to support their thesis. Considering the structure and the dynamics of the brain, they postulated that the human mind works as follows: The nuclear spin ensembles (NSE) in both neural membranes and proteins quantum mechanically process consciousness-related information such that conscious experience emerges from the collapses of entangled quantum states of NSE under the influence of the underlying spacetime dynamics. Said information is communicated to NSE through strong spin-spin couplings by biologically available unpaired electronic spins such as those carried by rapidly diffusing oxygen molecules and neural transmitter nitric oxides that extract information from their diffusing pathways in the brain. In turn, the dynamics of NSE has effects through spin chemistry on the classical neural activities such as action potentials and receptor functions thus influencing the classical neural networks of brain [6] The authors also gave some supporting evidence to such a formulation introducing indications for experimental verifications.

Also recently, [7], we have given direct formulation for a possible quantum mechanical model of consciousness based on the central role of the spin.

There is still another important reason to discuss here the real role of spin in biological dynamics.

The four bases in RNA sequences C, G, A and U ( or T in DNA) may be formalized by using Pauli matrices. Note that we do not speak here of some physical feature of such molecules but of their intrinsic representation and description. Of course, the four bases of RNA (or DNA) pertain to the most universal language and description of our biological reality. Let us explain it .Bases of the same heterocyclic kind (purine or pyrimidine) have the same signs .A proper reference frame may be introduced which slides along the RNA chain from 5' to the 3' end. Let the j-th and k-th nucleotides be paired. The base pair $X_j Y_k (j < k)$ is encountered twice :when the reference frame reaches position $j$, and from this position, nucleotide $X_j$ looks upright, but nucleotide $Y_k$ is upside-down; still, when reference frame reaches position $k$, from here nucleotide $Y_k$ looks upright, but nucleotide $X_j$ is upside-down. It follows that one may distinguish four base pair states of RNA since AU and UA, CG and GC, are no longer identical. $A_\downarrow$, $U_\downarrow$, $C_\uparrow$, $G_\uparrow$, $U_\uparrow$, $A_\uparrow$, $G_\downarrow$, $C_\downarrow$, once again, may be represented by Pauli matrices that in physics are representative of the spin but here represent base-pair values and still the base-pair creation and base-pair disruption. These results were obtained by Y. Magarshak [8]. We retain that they confirm fully our thesis. The notion of spin must be intended according to a very general meaning, that one of an entity that is articulated at a physical but also informative, and logic level until as a proto quantum like potential entity, whose great importance may be identified in biological as well as in physiological studies.

Let us consider still another important result that also legitimates the reason to consider the role of spin on a more general plane. Chaotic behaviors have been identified in a consistent number of signals pertaining to physiology and biology [9]. Starting with 1999, A. Jadczyk and R. Olkiewicz [10] showed that simultaneous measurements of non commuting spin components lead to a chaotic jump on a quantum spin sphere and to generation of specific fractal images on the basis of a non linear iterated function system. Thus, once again, non commutativity, as just was outlined also by M. Zak in a previous work [11], and spin may be also responsible of chaotic behaviors.

Several authors, [6,12] repeatedly evidenced that the spin is the essence of quantum mechanics having an ontological meaning. Hu and Wu repeatedly outlined that the driving force behind the evolution of Shrödinger equation is quantum spin and, since quantum entanglement arises from the evolution of Shrödinger equation the said spin is the genuine cause of quantum entanglement. To support this thesis we outline that recently we showed that Schrödinger equation is a manifestation of an abstract algebraic formulation in which the basic elements are given as well as in the Pauli spin formulation [13]. Quantum potentialities arise through quantum superposition principle that is admitted in Scrödinger equation.

## 4. On the Possibility to Introduce an Algebraic Structure as Quantum Like Scheme of Our Reality.

The conclusion of the previous section is that with the term spin we should intend an entity that seems to assume a general role in our reality for the variety of the dynamics that it is able to support and for the high differentiation of the processes to which it is able to oversee. We aim to give a reason for such an entity to oversee the natural phenomena at different levels. The reason could be that the so called spin as admitted in physics is really expression of a more general and differentiated essence and modality of self-fulfilment of reality at its various levels of manifestation. The confirmation could arise under a mathematical profile. It is known that on October 16 of 1843, a mathematician, Sir W.R. Hamilton, discovered hypercomplex numbers [14] that he initially identified as the algebra of pure time. The scientific community acknowledged lukewarmly such a new mathematical discovery . J.T. Graves stated "I have not yet any clear views on the extent to which we are at liberty to arbitrarily create new imaginaries and to endow them with supernatural properties"-such as non commutativity. In the $19^{th}$ century W.M. Clifford [15] completed the work initiated by Grassman and Hamilton giving a complete formulation of such algebraic structures. A rather trivial but interesting feature is that the Hamilton algebra may be also represented by matrices and in this case we re-find Pauli matrices and the non commutativity of the basic generators of the algebra in the same manner in which they appear in quantum theory of spin. Therefore , the reason for what we have previously called the universality of spin could be explained in the fact that its mathematical counterpart regards an algebraic structure, and algebraic structures arise in the description of natural processes and they have universal character. If we identify the spin Pauli matrices of physics in the inner body of an algebraic structure, in some sense we may attempt to show that such a structure represents a rough scheme of quantum like mechanics. If so, the universality of the algebraic structure should draw directly on the possibility to retain the same quantum like expressed theory as not specialized only at the quantum microphysical level for which it was introduced in 1927. In this manner we return to consider the problem of the potentiality and actuality that constitutes the basic aim of the present paper. As outlined in the first section, quantum mechanics exhibits two basic and original features. The first is that it admits potential as well as actualized states of physical reality. The second point is that it admits that, under suitable circumstances, we have a stochastic transition from potentiality to actualization of states via the so called unknown mechanism of the wave function reduction or psi collapse. In substance, if such a strong link exists between the given algebraic structure and a rough scheme of quantum mechanics, we must re-find in the algebra the results of quantum mechanics. First of all we have to delineate in detail the basic features of such an algebraic structure, discussing in particular its basic assumptions and the manner in which this algebra may derived on the basis of its starting axiomatic points. Shown that the introduced algebraic structure, represents actually a quantum like scheme of quantum mechanics, we may attempt to take a great step forward and this is to say to give for a first time a rigorous mathematical proof of the transition from potentiality to actualization that represents the basic indemonstrable fixed focus of all the quantum mechanics. We could be entirely successful since the algebraic structure that we will use could represent a more general ontological construction respect to a more restricted realization that could be represented from traditional quantum theory. These are the objectives that are reached in the following section. Here we will utilize the great work that, starting with 1981, was developed by Y. Ilamed and N. Salingaros [16]. We will follow the same technique that these authors used in their work. We anticipate here that only two basic assumptions, quoted as (a) and (b) in the following section, seem that are required in order to formulate a rough scheme of quantum mechanics.

**5. The Proof of Some Theorems.**

In this section we give a rigorous proof of theorems characterizing the algebra that we employ. We will follow some basic results that were previously given by Y. Ilamed and N. Salingaros [16] in 1981, when these authors studied in detail the algebra with three anticommuting elements.

Let us consider three abstract basic elements, $e_i$, with $i = 1,2,3$, and the element $e_0$, and let us admit the following two assumptions:

   a) it exists the scalar square for each basic element:
   $$e_1 e_1 = k_1 \ , \ e_2 e_2 = k_2, \ e_3 e_3 = k_3 \ \text{ with } \ k_i \in \Re \ . \qquad (1)$$
   In particular we have also that
   $$e_0 e_0 = 1.$$

   b) The basic elements $e_i$ are anticommuting elements, that is to say:
   $$e_1 e_2 = -e_2 e_1 \ , \ e_2 e_3 = -e_3 e_2, \ e_3 e_1 = -e_1 e_3. \qquad (2)$$
   In particular it is
   $$e_i e_0 = e_0 e_i = e_i \ .$$

Note that, owing to the axioms (a) and (b), the given basic elements must be considered abstract potential entities having the potentiality to simultaneously assume the numerical values $\pm k_i^{1/2}$. This is confirmed in particular by examining the (14) that is direct emanation of the two starting axioms. According to [16], these are the necessary and the sufficient conditions to derive all the basic features of the algebra that we employ. To give proof, let us consider the general multiplication of the three basic elements $e_1, e_2, e_3$, using scalar coefficients $\omega_k, \lambda_k, \gamma_k$ pertaining to some field:

$$e_1 e_2 = \omega_1 e_1 + \omega_2 e_2 + \omega_3 e_3 \ ; \ e_2 e_3 = \lambda_1 e_1 + \lambda_2 e_2 + \lambda_3 e_3 \ ; \ e_3 e_1 = \gamma_1 e_1 + \gamma_2 e_2 + \gamma_3 e_3. \qquad (3)$$

Let us introduce left and right alternation:

$$e_1 e_1 e_2 = (e_1 e_1) e_2 \ ; \ e_1 e_2 e_2 = e_1 (e_2 e_2) \ ; \ e_2 e_2 e_3 = (e_2 e_2) e_3 \ ; \ e_2 e_3 e_3 = e_2 (e_3 e_3) \ ; \ e_3 e_3 e_1 = (e_3 e_3) e_1 \ ;$$
$$e_3 e_1 e_1 = e_3 (e_1 e_1). \qquad (4)$$

Using the (4) in the (3) it is obtained that

$$k_1 e_2 = \omega_1 k_1 + \omega_2 e_1 e_2 + \omega_3 e_1 e_3;$$
$$k_2 e_1 = \omega_1 e_1 e_2 + \omega_2 k_2 + \omega_3 e_3 e_2;$$
$$k_2 e_3 = \lambda_1 e_2 e_1 + \lambda_2 k_2 + \lambda_3 e_2 e_3;$$
$$k_3 e_2 = \lambda_1 e_1 e_3 + \lambda_2 e_2 e_3 + \lambda_3 k_3;$$
$$k_3 e_1 = \gamma_1 e_3 e_1 + \gamma_2 e_3 e_2 + \gamma_3 k_3;$$
$$k_1 e_3 = \gamma_1 k_1 + \gamma_2 e_2 e_1 + \gamma_3 e_3 e_1 \ . \qquad (5)$$

From the (5), using the assumption (b), we obtain that

$$\frac{\omega_1}{k_2} e_1 e_2 + \omega_2 - \frac{\omega_3}{k_2} e_2 e_3 = \frac{\gamma_1}{k_3} e_3 e_1 - \frac{\gamma_2}{k_3} e_2 e_3 + \gamma_3 \ ;$$

$$\omega_1 + \frac{\omega_2}{k_1} e_1 e_2 - \frac{\omega_3}{k_1} e_3 e_1 = -\frac{\lambda_1}{k_3} e_3 e_1 + \frac{\lambda_2}{k_3} e_2 e_3 + \lambda_3 \ ;$$

$$\gamma_1 - \frac{\gamma_2}{k_1}e_1e_2 + \frac{\gamma_3}{k_1}e_3e_1 = -\frac{\lambda_1}{k_2}e_1e_2 + \lambda_2 + \frac{\lambda_3}{k_2}e_2e_3 \qquad (6).$$

For the principle of identity, we have that it must be

$$\omega_1 = \omega_2 = \lambda_2 = \lambda_3 = \gamma_1 = \gamma_3 = 0 \qquad (7)$$

and
$$\begin{aligned}-\lambda_1 k_1 + \gamma_2 k_2 &= 0 \\ \gamma_2 k_2 - \omega_3 k_3 &= 0 \\ \lambda_1 k_1 - \omega_3 k_3 &= 0\end{aligned} \qquad (8)$$

The (8) is an homogeneous system admitting non trivial solutions since its determinant $\Lambda = 0$, and the following set of solutions is given:

$$k_1 = -\gamma_2 \omega_3, \; k_2 = -\lambda_1 \omega_3, \; k_3 = -\lambda_1 \gamma_2 \qquad (9).$$

Admitting $k_1 = k_2 = k_3 = +1$, it is obtained that

$$\omega_3 = \lambda_1 = \gamma_2 = i \qquad (10)$$

Using the (3), the theorem is proven, showing that the basic features of the considered algebra are given in the following manner

$$e_1 e_2 = -e_2 e_1 = ie_3 \; ; \; e_2 e_3 = -e_3 e_2 = ie_1 ; \; e_3 e_1 = -e_1 e_3 = ie_2 \; ; i = e_1 e_2 e_3 \qquad (11).$$

The content of theorem 1. is thus established: given three abstract basic elements as defined in (a) and (b), an algebraic structure is established with four generators ($e_0, e_1, e_2, e_3$).

Note that the (11) represents one of the most basic relations in quantum mechanics. It has been here derived only on the basis of two algebraic assumptions, given respectively in (a) and (b).

We may now add some comments to the previous formulation.

The activity of scientific knowledge in two hundred years of development unequivocally shows that some algebraic structures arise naturally in the description of natural entities and phenomena. Thus, it is quite natural to attempt to identify the phenomenological counterpart of the algebraic structure given in (11). From (1) we have that

$$e_1^2 = 1, \; e_2^2 = 1, \; e_3^2 = 1 \qquad (12)$$

The (12) evidences that, being the $e_i$ abstract potential entities, we may choice to attribute them the numerical values of $\pm 1$. Admitting to be $p_1(+1)$ the probability to attribute the value $+1$ to $e_1$ and $p_1(-1)$ that one for $-1$, considering the corresponding notation for the two remaining basic elements, we may introduce the following mean values:

$<e_1> = (+1)p_1(+1) + (-1)p_1(-1)$ , $<e_2> = (+1)p_2(+1) + (-1)p_2(-1),$
$<e_3> = (+1)p_3(+1) + (-1)p_3(-1).$ (13)

It has been shown elsewhere [17] that

$$<e_1>^2 + <e_2>^2 + <e_3>^2 \leq 1 \qquad (14).$$

Let us observe that the (14) may be considered to represent a general principle of ontic potentialities and, in particular, it indicates that we never can attribute simultaneously definite numerical values to two basic elements $e_i$. In conclusion, as seen by the axioms (a) and (b), by the (11), by the (13) and the (14), we have delineated a rough scheme of quantum like theory through an algebraic structure. In this algebraic scheme some principles of the basic theoretical framework result to be represented.

These principles are that the given algebraic structure reflects an intrinsic indetermination and an ontic potentiality for its abstract elements. This means that, in absence of a direct numerical attribution, such basic elements are symbols that act in the algebra as such symbols, having an intrinsic indetermination and an ontic potentiality. This is to say that, in absence of attribution of a given numerical value, the basic elements $e_i$ operate in the given algebraic structure preserving the potentiality to assume a direct, possible, numerical value at any stage of the algebraic operations. In addition, let us consider, as example, to be $<e_1>=1$ (attribution of $+1$ to $e_1$), the (14) unequivocally shows that both $e_2$ and $e_3$ remain in the superposition of potential states of $+1$ and $-1$. Let us explain this last point in detail. The algebraic structure given in (1), (2), and (11) admits idempotents. Let us consider two of such idempotents:

$$\psi_1 = \frac{1+e_3}{2} \quad \text{and} \quad \psi_2 = \frac{1-e_3}{2} \tag{15}$$

It is easy to verify that $\psi_1^2 = \psi_1$ and $\psi_2^2 = \psi_2$. Let us examine now the following algebraic relations:

$$e_3\psi_1 = \psi_1 e_3 = \psi_1 \tag{16}$$

$$e_3\psi_2 = \psi_2 e_3 = -\psi_2 \tag{17}$$

Similar relations hold in the case of $e_1$ or $e_2$. The relevant result is that the (16) establishes that the given algebraic structure, with reference to the idempotent $\psi_1$, attributes to $e_3$ the numerical value of $+1$ while the (17) establishes that, with reference to $\psi_2$, the numerical value of $-1$ is attributed to $e_3$.

The conclusion is very important. The conceptual counter part of the (16) and (17) is that we are in presence of a self-referential process. On the basis of such self-referential process, as given in (16) and in (17), this algebraic structure is able to attribute a precise numerical value to its basic elements. Each of the three basic elements is able to make a transition from the condition of pure potentiality to a condition of actuality, that is to say in mathematical terms from the pure symbolic representation of the given abstract elements to that one of a real number. Let us remember that, on the basis of the (14), this self-referential process may regard each time one and only one of the three basic elements. In brief, for the first time we are analyzing an algebraic structure that represents a rough quantum like scheme and that, at the same time, as repeatedly admitted also in usual quantum mechanics, evidences, on the basis of a self-referential process, that it is possible a transition from potentiality to actualization as we discussed it in the first section of this paper.

Note also the importance of the (16) and the (17) from the view point of the logic. Through the self-referential process given in (16) and (17), our algebra recovers two first principles of logic that are the Principle of non-Contradiction and the Principle of the excluded Middle.

Obviously, in order to reach a rigorous formulation of such matter, the central question that mathematically arises is that we must give proof that it does exist and it may be carefully defined an algebraic structure that initially is given as by the (1), (2), and (11) and then it is characterized by the numerical attribution to one of its basic elements, as example to $e_3$, of one numerical value, say of $+1$ or of $-1$. The same conclusion holds if we consider a numerical attribution to $e_1$ or to $e_2$.

Let us consider the following argument.
If
$$e_3 \to +1 \tag{18}$$
we should have that
$$\psi_1 \to +1, \psi_2 \to 0, \tag{19}$$
and, in the (11),
$$e_1 e_2 = i, e_2 e_1 = -i, e_2 i = -e_1, ie_2 = e_1, e_1 i = e_2, ie_1 = -e_2 \tag{20}.$$

In other terms, if we attribute the numerical value of $+1$ to $e_3$ a new algebraic structure arises with new generators whose rules are given in (20) instead of in (11). Therefore, the arising central problem is to proof the real existence of such new algebraic structure. Note that, in the case of the starting algebraic structure we showed that it exists in the following manner

$e_1^2 = 1, e_2^2 = 1, e_3^2 = 1, i = e_1 e_2 e_3, e_1 e_2 = -e_2 e_1 = ie_3, e_2 e_3 = -e_3 e_2 = ie_1, e_3 e_1 = -e_1 e_3 = ie_2$.

In the present case, with $e_3 \to +1$, we have to show that it exists in the following manner

$e_1^2 = 1, e_2^2 = 1, i^2 = -1, e_1 e_2 = i, e_2 e_1 = -i, e_2 i = -e_1, ie_2 = e_1, e_1 i = e_2, ie_1 = -e_2$ (22).

We arrive at the proof of theorem 2: given the algebraic structure A, fixed as in the (1), (2), and (11), it exists an algebraic structure B, that we call a subalgebra of A, with basic elements (generators) given in (22). To proof, consider that we now attribute to $e_3$ the numerical value of $+1$ and so it is dismissed from the basic scheme of the three anticommuting basic elements. It is now replaced by $i$. Rewriting the (3) and performing calculations we arrive to the solutions of the (8) that are given in the following manner:

$k_1 = -\gamma_2 \omega_3, k_2 = -\lambda_1 \omega_3, k_3 = -\lambda_1 \gamma_2$ (22)

where this time it must be $k_1 = k_2 = +1$ and $k_3 = -1$. The solutions are given for

$\omega_3 = +1, \lambda_1 = -1, \gamma_2 = -1$ (23)

and consequently the (22) are proven as expected. Therefore it is shown that in the case $e_3 \to +1$, the subalgebra B exists having the basic features given in (22).

The theorem 2 may be shown also in the case in which we attribute to $e_3$ the numerical value of $-1$. We have

$e_3 \to -1$ (24)

and

$\psi_1 \to 0, \psi_2 \to -1$ (25)

and the subalgebra B is given in the following terms:

$e_1^2 = 1, e_2^2 = 1, i^2 = -1, e_1 e_2 = -i, e_2 e_1 = i, e_2 i = e_1, ie_2 = -e_1, e_1 i = -e_2, ie_1 = e_2$ (26)

The solutions of the (22) are given in this case by $\omega_3 = -1, \lambda_1 = +1, \gamma_2 = +1$. The theorem is shown also in this case. In a similar way it is obtained the proof when considering the case of attribution of a numerical value to $e_1$ or to $e_2$.

In this manner we have reached the central aim of the paper. Also if using an algebraic structure, this is the first time in which we are able to show the manner in which it is realized the passage from potentiality to actualization and it has been demonstrated by using a rigorous formulation based on two mathematical theorems. Since, as previously said, the counterpart exists in natural processes of the algebraic structures arising during their description, we expect that the two theorems demonstrate the passage from potentiality to actualization in our reality.

## 6. An Application in $\psi$ - collapse of Quantum Mechanics.

It is well known that one of the basic unsolved problems of quantum mechanics resides in the so called process of reduction of wave function or $\psi$-collapse.

Consider a two state quantum system S with connected quantum observable $\sigma_3$. It is known that we have

$$\psi = c_1 \varphi_1 + c_2 \varphi_2 \quad \text{with} \quad \varphi_1 = \begin{pmatrix} 1 \\ 0 \end{pmatrix} \quad \text{and} \quad \varphi_2 = \begin{pmatrix} 0 \\ 1 \end{pmatrix} \quad (27)$$

and

$|c_1|^2 + |c_2|^2 = 1$ (28)

It is still known that we may represent the state of such system by a density matrix $\rho$ given in the following terms

$$\rho = a + be_1 + ce_2 + de_3 \qquad (29)$$

with

$$a = \frac{|c_1|^2}{2} + \frac{|c_2|^2}{2}, \quad b = \frac{c_1^*c_2 + c_1c_2^*}{2}, \quad c = \frac{i(c_1c_2^* - c_1^*c_2)}{2}, \quad d = \frac{|c_1|^2 - |c_2|^2}{2} \qquad (30)$$

where in matrix notation, $e_1, e_2$, and $e_3$ are the well known Pauli matrices

$$e_1 = \begin{pmatrix} 0 & 1 \\ 1 & 0 \end{pmatrix}, \quad e_2 = \begin{pmatrix} 0 & -i \\ i & 0 \end{pmatrix}, \quad e_3 = \begin{pmatrix} 1 & 0 \\ 0 & -1 \end{pmatrix} \qquad (31)$$

It is also easily verified and well known that we may find a $2 \times 2$ matrix representation of the algebra A (as well as of the subalgebra B), given in the previous section, by using the same matrix configuration, given in (31). In conclusion we may write the (30) in explicit form in one of the two equivalent forms:

$$\rho = \frac{1}{2}(|c_1|^2 + |c_2|^2) + \frac{1}{2}(c_1c_2^*)(e_1 + e_2i) + \frac{1}{2}(c_1^*c_2)(e_1 - ie_2) + \frac{1}{2}(|c_1|^2 - |c_2|^2)e_3 \qquad (32)$$

or

$$\rho = \frac{1}{2}(|c_1|^2 + |c_2|^2) + \frac{1}{2}(c_1c_2^*)(e_1 + ie_2) + \frac{1}{2}(c_1^*c_2)(e_1 - e_2i) + \frac{1}{2}(|c_1|^2 - |c_2|^2)e_3 \qquad (33)$$

Let us admit now that we make a measurement of $\sigma_3$ with result +1. Admitting in this case that the subalgebra B obtained in the previous section is valid, we have that the (18), the (19), and the (22) are valid in the (32), and thus we have that the quantum interference terms disappear and the matrix density is reduced to

$$\rho_M = |c_1|^2 \times I \qquad I = unity matrix. \qquad (34)$$

In the case of the measurement of $\sigma_3$ is performed with result -1, the same subalgebra holds where now the (24), the (25), and the (26) are valid. Applied to the (33), still the interference terms disappear, and they give this time that

$$\rho_M = |c_2|^2 \times I \qquad (35)$$

As expected, the subalgebra B, introduced in the previous section, describes the collapse of the wave function in quantum mechanics.

**7. On the possibility for Quantum Ontic Potentiality to Explain a Central Role in the Dynamics of Living Matter.**

In the present version of this paper we would add to the mathematical proof, given in the previous two sections, a deepening on the central role that quantum ontic potentialities could have in the dynamics of the living matter. We will follow in detail an excellent paper that was recently written by P.C.W. Davies [18] on this subject. According to this author we start outlining that biological systems must be essentially intended as information processors. It must be outlined here that ideas similar to those given by P.C.W. Davies, were also introduced by A.Yu. Khrennikov [18] as well as also such author considered the notion of Transformers of Information when speaking of living systems. The basic key here is that if we reason in the framework of a classical context, the biological molecules must be intended as processors of bits of information. This is in a classical physical framework, while instead, in a quantum like context, biological molecules become processors of qubits of information, that is to say this time that the information processing becomes direct consequence of the superposition principle of quantum mechanics that, as previously discussed in detail, regulates ontic potential states in the dynamics of our reality. To be clear, it does not exist at all a proven verification that quantum mechanics operates at the level of biological matter attending such theory to the dynamics of the biological sphere with its fundamental and ontological features as in particular the superpositions and thus the ontological potentiality of states and the transition from potential to actualized states. However, repeating what it was just evidenced in [18], quantum mechanics works in living matter in explaining, as example, the shapes of biomolecules, the specificity of proteins or the templating functions of nucleic acids. It is at work in determining diffusion rates or membrane properties or the strengths of molecular bounds, all results that of course result so determinant in understanding basic features of living matter. It should be still a matter of a restricted epistemic vision to accept that quantum mechanics is only matter of calculations at atomic, molecular and biomolecular levels at the biological level. As discussed in the previous sections, and as it has been evidenced in particular through the theorem 2, this theoretical elaboration is a theoretical body that includes so advanced and so radically new principles and views on our reality as in particular all that in an ontic vision is linked to the superposition of states, to potentiality and actualization and still to many other well known features that render this theory completely new and radically innovative in the complex phenomenology and ontology of the dynamics of the reality. Therefore, the basic interest becomes to ascertain if and how such features, so complex to be admitted at an ontic level of our reality, are they at work in dynamics of the biological being.

There are many directions we may consider wit this perspective. The first, also indicated in [18], is that one that one may define of the quantum mutations. In the framework of the previously considered quantum ontic potentialities, one evaluates that the genetic code could be considered as a quantum code in which the superposition of coding states would act leading to spontaneous errors in base pairing. This idea may be dated back to the discovery by Crick and Watson of the structure of the DNA and to the consequent possibility that mutations could occur as result of quantum fluctuations, that therefore should became the source of random biological information. Here the quantum mechanism that may be advocated is that one of the quantum tunnelling that represents the most essential and also the most surprising process of all quantum mechanics. Proton tunnelling could indeed alter the structure of nucleotide bases leading to incorrect pair bonding. The basic key here is that the quantum tunnelling is the most evident example of ontic superposition of potential states that we have in Nature. In analogy with the (27) that we introduced previously to represent a quantum mechanical two level system, but remaining on the general plane of discussion, we have in this last case that at any time the state of the particle may be described by a state function that is a linear superposition of two potential states, one corresponding to the condition that the particle has tunnelled and one corresponding instead to the condition that the particle still has not tunnelled. We have that, still see also the (27),

$$\psi = c_1 \varphi_{tunnelling\,still\,not\,happened} + c_2 \varphi_{tunneling\,happened} \qquad (36)$$

with

$p_1 = |c_1|^2 =$ probability for tunnelling not to happen

and

$p_2 = |c_2|^2 =$ probability for tunnelling to happen,

$p_1 + p_2 = 1$

The (36) represents an excellent example of quantum superposition of states. The clue with biological dynamics was given by J. McFadden and J. Al-Khalili [19]. These authors started observing that the central assumption that mutations happen randomly has been challenged by the process called adaptive or directed mutation. It was also detected experimentally when a non-fermeating strain of Escheria coli was plated onto rich media containing lactose [19]. In experiments performed by J. Cairns et al. [20] ,papillae of $lac^+$ lactose fermenting mutants arose over a period of several weeks yet mutations that did not confer any selective advantage did not appear during incubation [19, 20]. $lac^+$ mutants arose instead in experiments with much more frequencies in absence of lactose. Adaptive mutations were also observed in other experiments [21]. Adaptive mutations are different from standard mutations since they occur in cell that are not dividing or dividing rarely, they are time dependent but not replication dependent, they appear only after the cell is exposed to selective pressure [19]. In order to explain adaptive mutations, different authors [22] proposed to consider that they could be generated by environment induced collapse of the wave function describing DNA in a superposition of mutational states. This is exactly the superposition of potential states and its actualization that in the previous section we have shown to be possible by the theorems 1, and 2. In [19] it was investigated the possibility discussed in [22] and a specific model of mutational process involving quantum tunnelling and the time of decoherence were specifically introduced. The authors considered the so called Lowdin two step model [23] for generation of mutations initiated by the proton tunnelling of an H-bonded proton between adjacent sites within base pairs. Considering DNA replication, the state will evolve to incorporate both the correct base C for tunnelling not happened and the incorrect base T for tunnelling happened. We will have a superposition of potential states, one consisting of the unmutated condition and the other consisting instead of the mutated condition. The daughter DNA will be a superposition of potentialities, mutated and unmutated conditions:

$$\psi = c_1 \varphi_{tunnelingnothappened} C_{state} + c_2 \varphi_{tunnelinghappened} T_{state} \qquad (37)$$

Again here the basic importance of the theorem 2 is clearly evidenced. In biological matter forms of potentialities coexist in order to give final forms of actualization.

Assumed the mutant form of the protein, as example
*lacZ*
containing an arginine→ histidine amino acid substation resulting in $lac^- \rightarrow lac^+$ mutation in the cell in absence of lactose, the final state of the cell will be still a superposition of unmutated and mutated conditions

$$\psi = c_1 \varphi_{tunnelingnothappened} C_{state} Arg._{state} + c_2 \varphi_{tunnelinghappened} T_{state} His._{state} \qquad (38)$$

that will actualize to a final form according to the theorem 2.

Quite similar is the important chapter on enzyme reactions. Still according to [18], enzymes are catalyzing proteins during biochemical reactions. They give so high reactions rates that we cannot expect to be explained in the framework of the usual catalytic mechanics. Also here quantum tunnelling explains an essential role and this decisive role has been evidenced by a series of detailed and important results [24]. Again here, the invoked tunnelling mechanism is expressed by the (36) and thus also in this relevant field of biological matter the importance to recall the concept of superposition of potentialities and then of actualization in biological dynamics, results to be decisive as stated by theorem 2.

Another way to consider quantum effects is in the sphere of the nanostructures [18]. The proton pump has the role to maintain the appropriate voltage across the cell membranes [25]. These structures are complex enzymes whose operation seems to be described by quantum one-dimensional nanotubes. Also membranes are involved in very complex process. Here it is expected that the quantization of nonlinear membrane vibrations in cells should exhibit quantum behaviours as a Bose condensate [18].

Finally, we have the great role that quantum mechanics could hold at the level of synaptic transmission among neurons. Again Eccles and Beck, as well as J. Walker, [26], argued that neuron firings should be regulated from quantum tunnelling, as in the (36), and J. Walker, in particular, gave theoretical results in accord also with the experimental data in relation to the $f_{mepp}$ of vesicle release. Recently we formulated some models legitimating the same advent of consciousness [7,17] on the basis of quantum tunnelling.

There is still another important question resulting very convincing in admitting a basic role of quantum superposition principle and quantum like general scheme in dynamics of living matter. It relates the genetic code. In section three we inferred about a quantum like scheme of DNA and RNA considering the results of Magarshak [6] and identifying base pair values and base-pair construction and destruction by spin Pauli matrices. Again, we consider here that the origin of the genetic code may be linked to quantum information processing. It has been outlined by Patel in

2001 [27] that the nucleotide bases could be represented and remain in a superposition of quantum states for the time necessary to participate to replication process. One may consider that the universal genetic code is based on triplets of nucleotides of four varieties that code for 20 or 21 amino acids. In 1996 [27] L.K. Grover found an optimal quantum search algorithm in the sphere of quantum computing. Indicating by N the number of objects that can be distinguished by a number of yes/no queries, this author found that

$$(2Q+1)arcsen(\frac{1}{\sqrt{N}}) = \frac{\pi}{2} \qquad (39)$$

Note that this is the result of quantum algorithm. This is to say that it does not use classical arguments as well as it does not use advanced quantum mechanics. It employs only the principle of superposition of quantum states and of quantum interference. The algorithm starts in fact assuming an uniform superposition of all the possible states corresponding to equal probability for every building block to be selected. Now, as stated in [26,18], the very convincing argument is that the (39) admits just the following solutions:

$Q = 1$ and $N = 4$;
$Q = 3$ and $N = 20.2$;

this is to say that such quantum algorithm, based on the quantum superposition principles, gives actually as solution for the genetic code the triplets of four varieties of 20-21 amino acids. If not due to unexpected trivial coincidence, this is a result that clearly evidences the presence of quantum mechanics at the basis of living matter.

It may be now also clear the reason because the reality of living matter would choice to operate on the basis of quantum processing (qubit) instead of classical processing (bit), on the basis of superposition of ontic quantum potentialities and consequent actualization instead of concrete actualizations only. The reason is in a kind of exaltation of its power of processing information in addition to more specific ontologic motivations. Correctly P.C. W. Davies arranges the problem [18]: the life and its origin are in some sense a kind of search problem. The subset of living systems is an extremely small fraction of the total space of complex systems. As example, the fraction of peptide chains having biological relevance is exponentially very small respect to the whole set of all the possible sequences. Only a very small fraction of all the nucleotide sequences code for the biological function. So the central arising question is the following: given a mixture of classical molecular building blocks, how may and, as P. Davies outlines [18], how did matter to find the appropriate and extremely improbable combination by change among all the possible combinations and in a contained period of time? Any classical explanation shows that it would take a time much longer than the age of the universe [18]. It is quantum mechanics to give the possibility to answer correctly since it, according to theorem 2, runs about potentialities and actualization. Let us look at the time evolution model that we roughly introduced in section two:

$$\psi_{n+1,A} + \varphi_{n+1,P}X_P = (\psi_{n,A} + \varphi_{n,P}X_P)(\psi_{n,A} + \varphi_{n,P}X_P) = \psi_{n,A}^2 + \varphi_{n,P}^2 + 2\psi_{n,A}\varphi_{n,A}X_P \quad (40)$$

At any stage of such supposed evolution this system processes simultaneously three information ways, it drags three possible values of actualization respect to only one that should compete in the classical case. Since quantum systems admit superposition of states, they may be hold and they may also search for a great variety of alternatives at the same time. As currently said, they may explore and process information in parallel instead of in series. In this manner such systems may explore a vast array of alternatives simultaneously and at a more accentuated speed respect to the traditional case. This is the reason because sequences of biological relevance may be found much faster than one may have in a classical non quantum framework.

We would now to face the last argument. The fractal nature of many biological systems is receiving considerable attention from some years. The basic feature of self-similarity has been identified in a lot of cases as in spatial structures or in temporal fluctuations of many biological signals as well as in ion channel kinetics, in fetal breathing, in human cognition, in neuron firings, in cardiovascular system. As known, the greatest part of such biological systems exhibit fluctuations like $1/f$ noise.

A number of different mechanisms have been proposed in order to explain the origin of such $1/f$ behaviour in nature and in biological matter. It has been argued that this behaviour could be merely a result of the multiple system inputs to the considered system [28]. Owing to the widespread nature of $1/f$ noise in different biological systems the $1/f$ behaviour could be due to the fact that the final output should be affected by many processes that act at different time scales and in fact it has been shown that some distributions of time scales lead to 1/f behaviour. Therefore, this could be the origin of the complex fluctuations and $1/f$ scaling so recurrently observed in biological systems. In brief, in many biological systems the output of a system may be seen as the result of different semiautonomous contributing systems operating at different time scales. A convincing example is that one of Heart Rate where regulation takes part by a beat-to-beat contribution deriving from the autonomic system. Here the heart rate oscillations experience the input of a great variety of distinct random processes over a great variety of different time scales. We have vagally- mediated random inputs with frequencies about ¼ sec. , baroreflex modulations at frequencies about 1/10 sec plus modulations due to the hormonal systems, posture, activity level, meals, sleep-wake cycle, circadian rhythm and still other contributions acting on the sinus node activity at different time scales. This is a model that explains $1/f$ behaviour. According to [28], the output model may be written at any time step, $k$, as

$$y(k) = \sum_{i=1}^{n} A_i x_i(k) \qquad (41)$$

where each input is assumed to be amplified by a constant $A_i$ representing the its relative effect on the output of the considered system [28]. Given this starting model one does not expect to find $1/f$ behaviour in the framework of a quantum mechanical scheme. To show this, let us consider first a classical scheme. Here the time fluctuations of the signal are considered to be due to a kind of a random relaxation process that has a time constant $\tau$ and it may be easily shown that the power spectral density goes following the general form

$$S(f) = \frac{g(\tau)}{1 + f^2 \tau^2} \qquad (42)$$

and the classical behaviour $constant/f$ or $constant/f^\beta$ may be easily identified.
In this case the probability for each input signal is expressed in the following manner
$$P(t) = e^{-t/\tau} \qquad (43)$$
and, as previously said, it corresponds to a perfect classical scheme. To proof this one may consider each arriving input as the result of an intrinsically unstable system that at each time has a definite probability to be present or not to be present. Considering two times $t_1$ and $t_2$ with $t_2 > t_1$ in this classical schema one has that
$$P(t_1)P(t_2) = P(t_1 + t_2) \qquad (44)$$
Since, by hypothesis, the probability for the input signal to be present depends only upon time, considering only statistically independent events, we have the (43) that in fact admits then the (43) as solution. However, if we substitute our starting schema, considering instead that we have this time a quantum mechanical mechanism generating the input instead of the previously admitted classical picture, we have an unstable quantum system as starting case and thus in this case the basic role of the superposition of the states (signal present and signal not present) must be essentially recalled. This question leads as consequence that we no more may adopt the (43) as probability. In this quantum case, instead of the (43), we have the following expression for probability:
$$P(t) = e^{-t^2/\tau^2} \qquad (45)$$
That, as it is easily seen, is profoundly different from the (43). In this case, however, the autocorrelation function will be

$$C(t) = A_i^2 e^{-t^2/\tau^2} \tag{46}$$

and the power spectrum will be

$$S(f) = Re\, A_i^2 \int_0^\infty e^{-2\pi i f t} e^{-t^2/\tau^2} dt \tag{47}$$

that still gives

$$S(f) \cong \frac{A_i^2 \tau}{1+(\pi f \tau)^2} \tag{48}$$

The power spectrum scaling exhibits a crossover from brown noise to white noise with crossover value given at $f^* = 1/\tau$. If only two processes will be superimposed in the sense of quantum mechanics we re find the $1/f^\beta$ behaviour of the power spectrum.

The important conclusion is therefore the following: also in the case of pure quantum mechanical contributions we find again the $1/f^\beta$ behaviour of power spectrum that is exhibited from the greatest variety of biological signals.

Finally, we have to mention here one of the most important experiments that in our opinion have been performed in the last few years on this field of research.

As we know, wave-particle duality relates a basic feature of quantum physical reality, in substance it regards the fact that a quantum object may exhibit either wave or particle properties depending on the experimental arrangement that we decide to use.

R. Feynmann repeatedly outlined that the strangeness of quantum mechanics may be seen analyzing in detail all that happens in the simple and elegant arrangement that we usually call the Two Slit Experiment. However, in spite of its strangeness, the wave-particle duality of massive objects is a true basic foundation of quantum physics. De Broglie's wave hypothesis, formulated in 1923, was also confirmed experimentally for atoms and molecules, as $He$ atoms and $H_2$ molecules, starting with 1932 by Estermann and Stern [29] by diffraction experiments. A renewed interest for molecular interferometry started in 1994 in consequence of the first observation of interference for $I_2$ by Ramsey and Bordé [30]. The reason to indicate here such experimental results is that we retain that a quantum like mechanics is also too much involved in the dynamics of biological matter, and a very recent result, obtained in the framework of molecular interferometry, confirms such an approach to matter. In a paper of 2003 the research group directed by Zellinger [31] reported to have obtained for the first time the demonstration of the wave and thus quantum nature of massive objects as biomacromolecules. Such quantum effects were observed for the tetraphenylporphyrin (TTP) and for the fluorinated fullerenes. The porphyrin structure is at the core of many complexes regarding biological matter as in particular biomolecules representing the color center in chlorophyll and in hemoglobin. On the other hand, the fluorofullerene is $C_{60}F_{48}$, thus it is a very massive object having 1632 amu. Dismissed or vanishing quantum effects should be expected for such molecules. Instead such results indicate that the de Broglie wave nature of objects seems that may be the matter of some a generalization developing a possible role also at the so high stage of the molecular scale. As discussed in the previous sections of the present paper, we retain that quantum mechanics may be generalized in order to be involved in the biological dynamics of macromolecules, and possibly to overcome the limits of scale at the elementary or at the atomic level only as actually it was in the initial vision of founding fathers of quantum theory. The theorem 2 that we have given in this paper, gives direct indication of the importance to have obtained a mathematical proof of the existence of potential states and of transition potentiality-actualization that represent the core of any quantum mechanical view on our reality.

## REFERENCES

1) A. Yu. Khrennikov, The principle of supplementarity: A contextual probabilistic viewpoint to complementarity, the interference of probabilities, and the incompatibility of variables in quantum mechanics. Foundations of Physics, 35, N. 10, 1655 - 1693 (2005).
   A. Yu. Khrennikov, Information dynamics in cognitive, psychological, social, and anomalous phenomena. Kluwer, Dordreht, 2004.
   A. Yu. Khrennikov, Nonlinear Schrödinger equations from prequantum classical statistical field theory, Physics Letters A, 357, N 3, 171-176 (2006).
   H. Atmanspacker, H.Primas, Epistemic and Ontic Quantum Realities, Foundations of Probability and Physics, edited by A. Khrennikov, Am. Institute of Physics, 2005, 49-61;
2) D. Aerts, S. Bundevoert, M. Czachor, B. D'Hooghe, L. Gabora, P. Polk, On the foundations of the theory of evolution, to appear on Systems Theory in Philosophy and Religion, Vols I, II, Windsor, Ontario, Canada: IIAS;
3) W. Pauli, Zeitschrift fur Physics, 80, 573, 1933
   P.A.M. Dirac, The Principles of Quantum Mechanics, Oxford,1935
   W. Pauli, V. Weiskoff, Helv. Physica Acta, 7, 709, 1934
4) P. O'Hara, Wave Particle Duality in General Relativity, arXiv:gr-qc/9701034 v1-15Jan 1997
5) P. Cordero, C. Teitelboim, Remarks on Supersymmetric Black Holes, Phys. Lett., 78B, 80-83, 1978
   R. Tabensky, C. Teitelboim, The Square Root of General Relativity, Phys. Lett., 69B, 453-456, 1977
6) H. Hu, M. Wu, Thinking outside the box: the essence and the implications of quantum entanglement, and references therein, see the site www.Quantumbrain.org for all the contributions of such authors;
7) E. Conte, G. Pierri, L. Mendolicchio, A. Federici, J.P. Zbilut, A quantum model of consciousness interfaced with a non-Lipschitz chaotic dynamics of neural activity, submitted to Chaos, Solitons and Fractals.
8) Y. Magarshak, Quaternion Representation of RNA Sequences and Tertiary Structures, BioSystems, 30, 21-29, 1993
9) J.B. Bassingthwaighte, L.S. Liebovitch, B.J. West, Fractal Physiology, Oxford University Press, 1994.
10) A. Jadczyk, On quantum iterated function systems, arXiv:nlin CD0312021 v2, 12 Mar. 2004 and references therein.
11) M. Zak, Dynamical Simulations of Probabilities, Chaos, Solitons and Fractals, 8, 5, 793-804, 1997;
12) D. Hestenes, Quantum mechanics from self interaction, Found. Phys.,15, 63-87, 1983;
    S. Esposito, On the role of spin in quantum mechanics, Found. Phys. Letters, 12, 165-171, 1999;
    G. Salesi, E. Recami, Hydrodynamics of spinning particles, Phys. Rev. A57, 98-105, 1998;
    I.R. Bogan, Spin: the classical to quantum connection, arXiv quant-ph/0212110 (2002).
13) E. Conte, A. Federici, A. Yu. Khrennikov, J.P. Zbilut, Is determinism the basic rule in dynamics of biological matter?, Quantum Theory: reconsideration of its foundations, Vaxjio University press, 2003, 639-675;
14) H. Hamilton, On a new species of imaginary quantities connected with a theory of quaternions, Royal Irish Academy, 2, 424-434, 1844
15) W. K. Clifford, Mathematical Papers, Edited by R. Tucker, London 1882
16) Y. Ilamed, N. Salingaros, Algebras with Three anticommuting Elements, J. Math. Phys., 22, 2091-2095, 1981, and N. Salingaros, Algebras with anticommuting elements II, J. Math. Phys. 22, 10, 2096-2100, 1981

17) E. Conte, A. Federici, L. Mendolicchio, G. Pierri. J.P Zbilut, On a model of the neuron with quantum mechanical properties, in press on Chaos, Solitons and Fractals.
18) P. C.W. Davies, Does quantum mechanics play a non trivial role in life?, BioSystems, 78, 69-79, 2004; A. Yu. Khrennikov, Classical and quantum mechanics in information spaces with applications to cognitive, psychological, social and anomalous phenomena, Found. of Phys., 29, 7, 1065-1098 and Information Dynamics in cognitive, psychological, social and anomalous phenomena, Klwver, Dordreht, 2004;
19) J. Mc. Fadden, J. Al-Khalili, A quantum mechanical model of adaptive mutation, BioSystems, 50, 203-211, 1999;
20) J. Cairns, J. Overbaugh, S. Milar, The origin of mutans, Nature, 335, 142-145, 1988;
21) see ref. 20 and B.G. Hall, Adaptive mutagenesis at ebgR is regulated by PhoPQ., J. Bacteriol. 180, 2862-2865, 1998 and references therein;
22) A. Goswami, D. Todd, Is there conscious choice in directed mutation, phenocopies, and related phenomena?, Physiol. Behav. Sci., 32, 132-142, 1997
V.V. Ogyzko, A quantum theoretical approach to the phenomenon of directed mutations in bacteria, BioSystems, 43, 83-95, 1997;
23) P.O. Lodwin, Quantum genetics and the aperiodic solid,….. , Advances in Quantum Chemistry, vol.2, Academic Press, New York, 213-360, 1965;
24) M. Garcia-Viloca, J. Gao, M.Karplus, D.G. Truhlar, How enzymes work: analysis by modern rate theory and computer simulations, Science, 303, 186-195, 2004;
25) D.N. Silvermann, Marcus rate theory applied to enzymatic proton transfer, Biochem. Biophys. Acta,1458, 8-19, 2000;
26) F. Beck, J. C. Eccles, Quantum aspects of brain activity and the role of consciousness, Proc. Nat. Acad. Sci. Usa, 89, 11357-11362, 1992;
E.H. Walker, Quantum tunnelling in synaptic and ephaptic transmission, Int. J. Quantum Chem.11, 103-127, 1977;
27) L.K. Grover, A fast quantum mechanical algorithm for database search, Proceedings of the 28[th] Annual ACM symposium on the theory of computing, 212, 1996;
28) J. M. Hausdorff, C-K Peng, Multiscaled randomness: A possible source of 1/f noise in biology, Phys. Rev. E, 54, 2, 2154-2156, 1996;
29) I. Estermann, Stern A., Z. für Physik, 61, 95-99,1930;
30) N. F. Ramsey, Molecular beams, Oxford University Press, 1985;
C.J. Bordé, N. Courtier, F.D. Burek, A. N. Goncharov, M. Gorlicki, Phys. Lett. A, 188,187-191, 1994;
31) H. Hauckermuller, S. Uttenthaler, K. Hornberger, E. Reiger, B. Brezger, A. Zeilinger, H. Arndt, The wave nature of biomolecules and fluorofullerenes, arXiv. Quant-ph/0309016 v1 1 Sept. 2003.